\documentclass[twocolumn,prb,superscriptaddress,amsmath,amssymb,aps]{revtex4}
\usepackage{amsmath}
\usepackage{bm}
\usepackage{braket}
\usepackage{color}
\usepackage{url}
\usepackage[latin1]{inputenc}
\usepackage{tikz}
\usepackage[toc,page]{appendix}
\makeatletter
\def\BState{\State\hskip-\ALG@thistlm}
\makeatother

\begin{document}

\title {Accelerating spin-space sampling by auxiliary spin-dynamics and temperature-dependent spin-cluster expansion}

\author{Ning~Wang}
\affiliation{ICAMS, Ruhr-Universit\"{a}t Bochum, Universit\"{a}tstr. 150, 44780 Bochum, Germany }

\author{Thomas~Hammerschmidt}
\affiliation{ICAMS, Ruhr-Universit\"{a}t Bochum, Universit\"{a}tstr. 150, 44780 Bochum, Germany }

\author{Jutta~Rogal}
\affiliation{ICAMS, Ruhr-Universit\"{a}t Bochum, Universit\"{a}tstr. 150, 44780 Bochum, Germany }

\author{Ralf~Drautz}
\affiliation{ICAMS, Ruhr-Universit\"{a}t Bochum, Universit\"{a}tstr. 150, 44780 Bochum, Germany }

\begin{abstract}
Atomistic simulations of thermodynamic properties of magnetic materials rely on an accurate modelling of magnetic interactions and an efficient sampling of the high-dimensional spin space. Recent years have seen significant progress 
with a clear trend from model systems to material specific simulations that are usually based on electronic-structure methods. 
Here we develop a Hamiltonian Monte Carlo framework that makes use of auxiliary spin-dynamics and an auxiliary effective model, the temperature-dependent spin-cluster expansion, in order to efficiently sample the spin space. 
Our method does not require a specific form of the model and is suitable for simulations based on electronic-structure methods. 
We demonstrate fast warm-up and a reasonably small dynamical critical exponent of our sampler for the classical Heisenberg model. 
We further present an application of our method to the magnetic phase transition in bcc iron using magnetic bond-order potentials. 
\end{abstract}


\maketitle

\section{Introduction}

Understanding and predicting thermodynamic properties of magnetic materials is not only of general scientific interest, but also crucial for many applications, 
e.g., magnetic refrigeration \cite{Vitalij1999}, iron and steel design \cite{Hickel2012}, and optimization of magnetic nanoparticles in biomedicine\cite{Lu2007}. 
In a spin system, the potential energy $E$ is defined as a function of the spin configuration $\{ \mathbf{s}_1, \mathbf{s}_2...\mathbf{s}_N \}$. Finite-temperature properties are then investigated using Boltzmann statistics,
\begin{equation}\label{thermal_average_spin_space}
 \langle  O \rangle = \frac{1}{Z} \int \prod_i d \mathbf{s}_i \, \pi(\{ \mathbf{s}_i \})  O(\{\mathbf{s}_i\}),
\end{equation}
with the configurational probability density in spin space,
\begin{equation}\label{configurational_dist}
 \pi(\{ \mathbf{s}_i \}) = \exp \left[-\frac{1}{k_\text{B} T} E(\{\mathbf{s}_i\}) \right] \,,
\end{equation}
where $k_\text{B}$ is the Boltzmann constant and $T$ the temperature. $O$ may be any spin-dependent observable, and $Z$ is the partition function in the classical limit and contains no longitudinal spin-fluctuations\cite{Dietermann2012, Ruban2007, Dudarev2012}. 

Frequently, model Hamiltonians are used for evaluating spin-dependent observables, such as the rigid-lattice Heisenberg model or its coordinate-dependent variants that account for spin-lattice coupling \cite{Ma2008}.  
For materials design it would be desirable to work with material specific Hamiltonians that explicitly take into account the electronic structure. This would also facilitate an adequate treatment of spin fluctuations in itinerant-electron magnets\cite{Moriya1985}, 
complex exchange-interactions in Fe\cite{Kvashnin2016} and magnon-phonon coupling in magnetic transition metals\cite{Fritz2014}. 
Recent years have seen significant progress in the development of electronic-structure based models, e.g., density functional theory for non-collinear magnetism \cite{Kubler1988, Stocks1998, Hobbs2000, Blugel2004, Peralta2007, Ma2015}, 
non-collinear magnetic tight-binding\cite{Mukherjee2001, Barreteau2015} or bond-order potentials\cite{ Drautz2011, Michael2014, Ford2015, Drautz2015}.

For model Hamiltonians there are many Monte Carlo (MC) sampling algorithms that efficiently sample the spin space, but unfortunately none of them is suitable for electronic-structure based models. We notice that there are three main differences between model Hamiltonians and electronic-structure based models.
First, model Hamiltonians, at least most of them, contain only pair-wise interactions, while electronic-structure based models require many-spin interactions. Second, the range of the interaction is different. In model Hamiltonians typically only 
first and second nearest-neighbour interactions are taken into account, 
while in principle all spins are coupled in electronic-structure based models. Third, electronic-structure based models are orders of magnitude slower in the evaluation of the Hamiltonian.
These differences prohibit application of many efficient MC algorithms. 
For example, the checkerboard MC algorithm \cite{Landau1990} is not applicable to electronic-structure based models as the system cannot be decomposed into non-interacting sublattices and the checkerboard-decomposition method is not applicable.  
The Swendsen-Wang \cite{Swendsen1987} and Wolff\cite{Wolff1989} cluster algorithms significantly reduce the correlation of samples, but they only work for models which may be mapped onto percolation models, which is difficult for electronic-structure based models.  
The over-relaxation\cite{Brown1987,Creutz1987} algorithm works well for the classical 
Heisenberg model, but its efficient implementation depends upon the checkerboard-decomposition method \cite{Landau2005}. The heat-bath spin dynamics\cite{Dudarev2011, Tranchida2018} might be seen as a variant of  the over-relaxation algorithm and suffers from the same problem. 
The Wang-Landau sampling\cite{Wang2001} can in principle overcome the critical slowing-down, but the convergence of the density of states usually requires millions of energy evaluations, which is not affordable for electronic-structure based methods. 
In a recent application of the Wang-Landau sampling to first-principle non-collinear magnetism 590,000 energy evaluations were performed yielding a reasonable estimate of the density of states\cite{Eisenbach2011}, 
but despite the huge computational effort the results could not be fully converged.

Our work builds on recent progress in the Hamiltonian Monte Carlo (HMC)\cite{Neal2011,Mark2011,beskos2013, Betancourt2017_2, Betancourt2017_1, Wang2013, Hoffman2014} and 
efficient methods to accelerate first-principles thermodynamic calculations\cite{Hellman2013, Grabowski2009, Duff2015}. For the former, a rigorous theoretical proof has been given that underpins the empirical success of HMC \cite{Betancourt2017_1} and 
the theoretical considerations \cite{beskos2013} and techniques\cite{Neal2011, Betancourt2017_2, Wang2013, Hoffman2014} developed for the automatic tuning of its hyper-parameters. In the latter, effective potentials are 
employed to speed up first-principles thermodynamic calculations, as the direct calculations with first-principles methods are too expensive. 
However, the conventional HMC algorithm is not applicable to spin systems as the spin length is not preserved in standard molecular dynamics that is used in HMC. The 
effective-potential methods in literature \cite{Hellman2013, Grabowski2009, Duff2015} are designed to accelerate calculations of free energies and cannot be applied to evaluate other thermodynamic quantities straightforwardly. 
These considerations form the basis for the methods developed in this work. First, we propose an auxiliary spin-dynamics as a basis for a HMC algorithm for spin systems that  
rigorously preserve spin lengths. Second, we propose a HMC framework in which the temperature-dependent spin-cluster expansion (SCE)\cite{Drautz2004, Drautz2005} is used as an auxiliary model to further accelerate the sampling of the spin space. 

The paper is structured as follows: 
we first introduce the HMC algorithm for spin systems using auxiliary spin-dynamics, and discuss the automatic tuning of its hyper-parameters. Then we 
introduce the temperature-dependent SCE as an auxiliary model to accelerate the sampling of the spin space.
In Sec. \ref{results_discussion}, we employ the classical Heisenberg model to demonstrate the efficiency of our method, and apply our algorithm to sample the magnetic phase transition in bcc iron with magnetic bond-order potentials\cite{Drautz2011, Matous2011, Ford2015}.

\section{Methodology}\label{methodology_section}
Our target is to draw efficiently independent samples according to the configurational probability density defined in spin space. To this end we extend HMC \cite{Neal2011,Betancourt2017_1} for the sampling of spin space variables. HMC does not sample the configurational distribution directly but a joint distribution of positions $q$ and momenta  $p$, 
\begin{equation} \label{joint_distribution}
 \pi(q, p) = \frac{1}{Q} \mathrm{exp}\left[ -\frac{1}{k_\text{B} T} \mathcal{H}(q,p) \right ],
\end{equation}
where $Q$ is the partition function in the phase space. The marginal distribution of $q$ then restores the target distribution. 

For sampling spin space using HMC we introduce auxiliary spin angular velocities $\bm{\omega}_i$ as canonical variables of the spin directions $\{ \bm{\omega}, \bm{s} \}$ in formal analogy to the classical canonical variables $\{ \bm{p}, \bm{q} \}$ and define the Hamiltonian as 
\begin{equation}\label{spin_Hamiltonian}
 \mathcal{H} = \frac{I}{2} \sum_i \bm{\omega}_i^T \cdot  \bm{\omega}_i + E(\{ \mathbf{s}_i\} ),
\end{equation}
where $I$ is a fictitious mass that later is used as a parameter to optimize the efficiency of the HMC sampling.

The configurational probability density in spin space, Eq. \eqref{configurational_dist}, is restored by the marginal distribution of the configurational variables in phase space,
\begin{equation} \label{marginal_distribution}
 \pi(\{\bm{s}_i\}) =  \int \prod_i d{\bm \omega}_i \, \pi(\{\bm{\omega}_i, \bm{s}_i\}),
\end{equation}
with the joint probability density defined as
\begin{equation}\label{joint_distribution}
\pi(\{\bm{\omega}_i, \bm{s}_i\}) = \frac{1}{Q} \mathrm{exp}\left[ -\frac{1}{k_\text{B} T} \mathcal{H}(\{\bm{\omega}_i,\bm{s}_i \}) \right ].
\end{equation}
We may thus sample the joint probability density Eq. \eqref{joint_distribution} and then obtain the configurational probability density from Eq. \eqref{marginal_distribution}, which leads to the correct thermal average Eq. \eqref{thermal_average_spin_space}. 
For sampling Eq. \eqref{joint_distribution}, we employ Hamiltonian dynamics together with Monte Carlo sampling, as outlined in the following.  

Different from usual spin dynamics, which is based on a first-order differential equation in time\cite{Antropov1996,Gilbert2004, Ma2008}
\begin{equation}\label{classical_spin_dynamics}
\frac{d \mathbf{s}_i}{d t} = \frac{\gamma_e}{m_i}  \frac{\partial E}{\partial \mathbf{s}_i}   \times \mathbf{s}_i,
\end{equation}
where $m_i$ is the magnitude of magnetic moments, and $\gamma_e$ is the gyromagnetic ratio for an electron spin, the auxiliary Hamiltonian, Eq. \eqref{spin_Hamiltonian}, dictates that the spins follow conventional Hamiltonian dynamics for rigid bodies
\begin{equation} \label{equations_of_motion}
\begin{aligned}
 & I\frac{d\bm{\omega}_i}{dt} = \frac{\partial E}{\partial \mathbf{s}_i} \times \mathbf{s}_i, \\
 & \frac{d\mathbf{s}_i}{dt} = \bm{\omega}_i \times \mathbf{s}_i.
\end{aligned}
\end{equation}
It is evident that the spin dynamics described by Eq. \eqref{classical_spin_dynamics} and Eq. \eqref{equations_of_motion} is different. Eq. \eqref{equations_of_motion} describes a completely fictitious dynamics of the spins that, however, 
by construction may be used to sample spin space according to the probability density  $\pi(\{\mathbf{s}_i\})$. We denote Eq. \eqref{classical_spin_dynamics} as the semi-classical spin dynamics and Eq. \eqref{equations_of_motion} as the auxiliary spin-dynamics in this paper. 

While it is very difficult or impossible to find an effective symplectic time-reversible integrator for the semi-classical spin dynamics \cite{McLachlan2014,KRECH19981}, this is not a problem for the auxiliary spin-dynamics. We combine the velocity Verlet method \cite{Swope1982} and a spin rotation scheme to obtain an efficient numerical integration 
algorithm that is time-reversible, area-preserving and preserves spin length.  
There are three steps per update, 
\begin{equation} \label{integration_scheme}
\begin{aligned}
 &  \bm{\tilde{\omega}}^{t+1}_i =  \bm{\omega}^t_i+ \frac{1}{2} \frac{\epsilon}{I} \cdot\frac{\partial E}{\partial \mathbf{s}^t_i} \times 
 \mathbf{s}^t_i,\\
 & \mathbf{s}^{t+1}_i = \mathbf{D}(\bm{\tilde{\omega}}^{t+1}_i, \epsilon) \cdot \mathbf{s}^t_i, \\
 &   \bm{\omega}^{t+1}_i =   \bm{\tilde{\omega}}^{t+1}_i + \frac{1}{2} \frac{\epsilon}{I} \cdot\frac{\partial E}{\partial \mathbf{s}^{t+1}_i} \times \mathbf{s}^{t+1}_i,
\end{aligned}
\end{equation}
where  the index $i$ denotes spin, $t$ the current state and $t+1$ the next state.   $\epsilon$ is the time step,
and $\mathbf{D}(\bm{\tilde{\omega}}^{t+1}_i, \epsilon)$ is a $3\times3$ rotation matrix \cite{Omelyan2001},
\begin{equation}
 \mathbf{D}(\bm{\tilde{\omega}}^{t+1}_i, \epsilon) = \mathbf{I} + \mathbf{W}^{t+1}_i\, \mathrm{sin}(\omega^{t+1}_i \epsilon) + 
 (\mathbf{W}^{t+1}_i)^2 \left [ 1-\mathrm{cos}(\omega^{t+1}_i \epsilon)  \right ].
\end{equation}
$\omega^{t+1}_i$ is the magnitude of $\bm{\tilde{\omega}}^{t+1}_i$, and $\mathbf{W}^{t+1}_i$ is a 
skew - symmetric matrix with $\mathbf{W}^{t+1}_{i,XY}=-\hat \omega^{t+1}_{i,Z}$, $\mathbf{W}^{t+1}_{i,XZ}=
\hat \omega^{t+1}_{i,Y}$, $\mathbf{W}^{t+1}_{i,YZ}=-\hat \omega^{t+1}_{i,X}$, where
${\hat{\bm{\omega}}}^{t+1}_{i}$ is the directional vector of $\bm{\tilde{\omega}}^{t+1}_{i}$. $X$, $Y$ and $Z$ denote 
Cartesian components.

We may now employ standard HMC for sampling the spin space probability density $\pi(\{\mathbf{s}_i\})$ on the basis of the auxiliary spin-dynamics. There are three steps per MC update. The first step performs a Gibbs sampling of angular velocities, in which we fix the configurational variables $\{ \mathbf{s}_i  \}$ and sample the angular velocities according to their conditional, Gaussian distribution. 
In the second step the auxiliary spin-dynamics is run for a specific trajectory of length $L$. In the third step the Metropolis-Hastings acceptance-rejection is performed for the proposal state generated by auxiliary spin-dynamics in order to guarantee detailed balance. The acceptance probability is given by 
\begin{equation}\label{accep_1}
 p^{\mathrm{acc}}(\mathbf{x}_{\mathrm{new}} | \mathbf{x}_{\mathrm{old}}) = \mathrm{min} \left \{ 1, \mathrm{exp} \left(-\frac{\Delta \mathcal{H}_{\mathrm{old}\rightarrow \mathrm{new}}}{k_{\text{B}}T}  \right)   \right \},
\end{equation}
where $\mathbf{x}_{\mathrm{old}}$ and $\mathbf{x}_{\mathrm{new}}$ are the state variables containing both the spins and the angular velocities. 
Since in Hamiltonian dynamics the energy is conserved, the Metropolis-Hastings acceptance criterion only corrects for changes in energy due to numerical integration errors, resulting in high acceptance ratios.
Repeating the three steps above leads to a Markov chain which obeys the joint distribution and is used to evaluate thermal averages. 

There are three hyper-parameters in our algorithm, the mass $I$, the time-step $\epsilon$ for numerical integration and the trajectory length $L$. 
They have no effect on accuracy of the MC sampling but strongly influence the efficiency and in practical applications should be set automatically without user intervention. We adapt the time step such that the exponential moving average of the acceptance probability is in the range from 0.6 to 0.7, which is centered around the optimal value 0.651 suggested by Beskos et al.\cite{beskos2013} More specifically, if the acceptance probability is smaller (larger) than 0.6 (0.7), 
the time step is decreased (increased) by a specific factor. For the tuning of the trajectory length $L$, we employ an empirical termination criterion, the U-turn termination\cite{Hoffman2014}, as given in Appendix~\ref{U_turn_termination}.
The hyper-parameters are only tuned in the warm-up phase and then fixed to leave the distribution function invariant in the sampling phase. We fix the mass $I$ to $1~\text{eV} \text{fs}^2$ in our work, which is an empirical optimal value according to our tests. 

We denote the HMC based on auxiliary spin-dynamics as Algorithm I in this paper. 
We next incorporate the temperature-dependent SCE into Algorithm I to further accelerate the sampling for expensive spin models and denote the new method Algorithm II. This is based on the observation that the auxiliary spin-dynamics is only used to generate the proposal state and may be run with a cheaper auxiliary model instead of 
the expensive target one. This leaves the sampling correct as long as the detailed balance is guaranteed for the target model through the Metropolis-Hastings acceptance-rejection. 
The criterion for the auxiliary model is that it should be as close to the target one as possible since the acceptance probability is now determined by the difference between auxiliary and target models,
\begin{equation} \label{accep_auxiliary}
\begin{aligned}
 & p^{\mathrm{acc}}(\mathbf{x}_{\mathrm{new}} | \mathbf{x}_{\mathrm{old}}) \\
 & =  \mathrm{min} \left[ 1, \mathrm{exp} \left(-\frac{\Delta \mathcal{H}^{\mathrm{target}}_{\mathrm{old}\rightarrow \mathrm{new}}}{k_{\text{B}}T}  \right)   \right]   \\
 & =   \mathrm{min}  \left [ 1,\mathrm{exp} \left(\frac{\Delta \mathcal{H}^{\mathrm{auxiliary}\rightarrow \mathrm{target}}_{\mathrm{old}}}{k_{\text{B}}T}  \right) \times \right. \\
 & \left.  \,\,\,\,\,\,\,\,\,\,\,\,\,\,\,\,\,\,\,\,\,\,\,\,\,\,\,\,\,\,  \mathrm{exp} \left(-\frac{\Delta \mathcal{H}^{\mathrm{auxiliary}}_{\mathrm{old}\rightarrow \mathrm{new}}}{k_{\text{B}}T}  \right) \times  \right. \\
 & \left.  \,\,\,\,\,\,\,\,\,\,\,\,\,\,\,\,\,\,\,\,\,\,\,\,\,\,\,\,\,\,   \mathrm{exp} \left(-\frac{\Delta \mathcal{H}_{\mathrm{new}}^{\mathrm{auxiliary}\rightarrow \mathrm{target}}}{k_{\text{B}}T}  \right)  \right] .
\end{aligned}
\end{equation}
In Eq. \eqref{accep_auxiliary}, the first and last exponentials contain the energy difference between the auxiliary and the target model for the old and new states, respectively. 
The second exponential arises from the numerical error of the integration of the auxiliary spin-dynamics 
with the auxiliary model, which is typically a small contribution. 

\begin{figure}[h!]
\centering
\includegraphics[width=0.9\columnwidth]{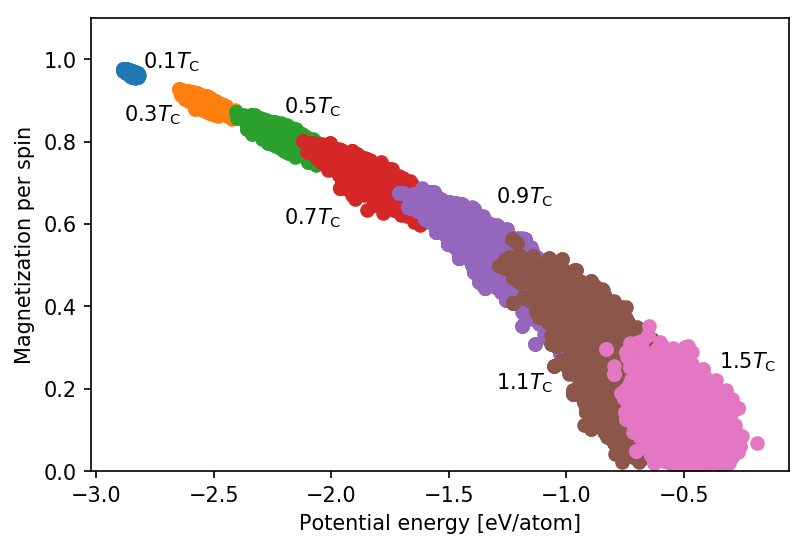}
 \caption{Magnetization-energy plot for spin configurations of a $6\times6\times6$ simple cubic 
 lattice at different temperatures. The classical ferromagnetic Heisenberg model is employed in this test. 
 $T_\mathrm{C}$ is its Curie temperature. The exchange parameter $J$ is chosen to be $1~\text{eV}$. }
 \label{area_exploration}
\end{figure}

For spin systems, an ideal auxiliary model is the spin-cluster expansion\cite{Drautz2004, Drautz2005} which may be fitted to accurately reproduce the target model. Here we propose to generate temperature-dependent SCEs due to two considerations. 
First, only a specific area in the spin space is explored with high probability at a given temperature, as shown in Fig. \ref{area_exploration}, 
where we plot the magnetization and the potential energy of spin configurations at different temperatures 
for a classical ferromagnetic Heisenberg model. Clearly, the configurations at different temperatures are clustered into specific areas. This locality makes it easier to fit a temperature-dependent effective potential. 
Second, the temperature dependence of the exchange parameters is inherited in some models, 
e.g., in spin-density-functional-theory for itinerant-electron magnets\cite{Rosengaard1997}. In practice, we collect the spin configurations in the warm-up phase and fit the temperature-dependent SCE at different temperatures separately. 
The extra cost required for fitting the temperature-dependent SCEs is normally negligible compared to the total computational time for sampling. The auxiliary spin-dynamics is then run with the temperature-dependent SCE to generate proposal states while the Metropolis-Hastings 
acceptance-rejection is performed for the target models in order to guarantee correct sampling. Implementation details are given in Appendix \ref{implementation_detail}.  
We note that no gradient calculations for the electronic-structure based models are required in this algorithm as the auxiliary spin-dynamics is run with the temperature-dependent SCE, 
which is another advantage of our method since the evaluation of gradients usually requires considerable extra computational cost.

\section{Results and Discussion}\label{results_discussion}
We demonstrate two applications in this section. 
In the first application, we perform efficiency tests for Algorithm I using the classical Heisenberg model. 
In the second application,  we test and discuss Algorithm II using magnetic bond-order potentials\cite{Drautz2011, Michael2014, Ford2015, Drautz2015}.   

\subsection{Application to the classical Heisenberg model} 
We first employ the classical ferromagnetic Heisenberg model,
\begin{equation}
 E = -J \sum_{<ij>} \mathbf{s}_i \cdot \mathbf{s}_j, 
\end{equation}
on a simple cubic lattice to perform an efficiency test for HMC, Algorithm I. $J$ is the exchange parameter, and $<ij>$ denotes the first-nearest-neighbour pairs without double counting. $J$ is chosen to be $1~\text{eV}$ in this test. 

\begin{figure}[h!]
\centering
\includegraphics[width=0.9\columnwidth]{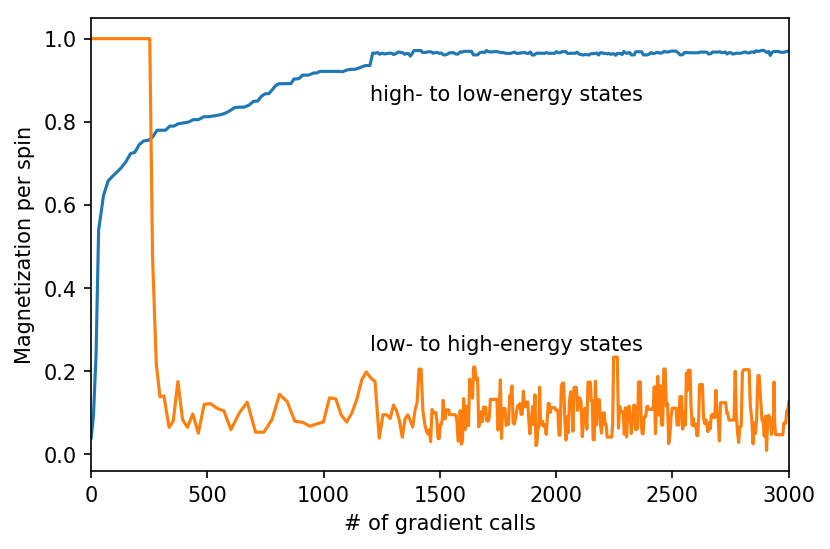}
 \caption{Evolution of the magnetization of a $6\times6\times6$ simple cubic lattice for the classical Heisenberg model in the warm-up phase. 
 Orange line: the initial state is the ferromagnetic ground state and the temperature is 2~$T_\mathrm{C}$. Blue line: the initial state is 
 a random-spin state and the temperature is 0.1~$T_\mathrm{C}$.  }
 \label{magnetization_warmup}
\end{figure}

In Fig. \ref{magnetization_warmup}, we measure the warm-up efficiency for a $6\times6\times6$ simple cubic lattice from the high (low)-energy to low (high)-energy states. 
As expected, the efficiency from low-energy to high-energy states is higher than the other way round, and the former (latter) takes around 400 (1200) gradient calls. 

\begin{figure}[h!]
\centering
\includegraphics[width=0.9\columnwidth]{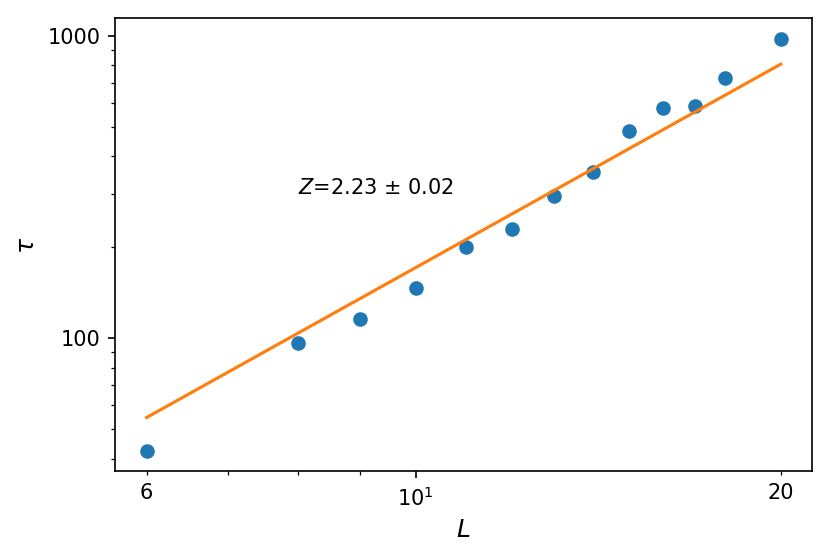}
 \caption{Log-log plot of the relaxation time for the magnetization in units of gradient calls. $L$ represents the side length of the cubic simulation cell and $\tau$ represents the relaxation time. The error bars of the relaxation time are smaller 
 than the symbol size. The classical Heisenberg model on the simple cubic lattice is used here.}
 \label{correlation_time_heisenberg}
\end{figure}

We next fix the temperature to the critical temperature of the 3D classical Heisenberg model ($T_\mathrm{C} = 1.4459~ J/k_{\mathrm{B}}$) \cite{Landau1990}, and estimate the dynamical critical exponent. The estimation is based on the 
dynamical finite-size scaling ansatz\cite{Suzuki1977}
\begin{equation}
 \tau \approx L^z,
\end{equation}
where $L$ is the side length of the cubic simulation cell. $\tau$ is the relaxation time of magnetization and evaluated according to
\begin{equation}
\phi(t) = A e^{-t/\tau},
\end{equation}
where $\phi(t)$ is the time auto-correlation function of magnetization. The relaxation time $\tau$ is measured in units of gradient calls instead of MC step as multiple gradient calls may be required for one MC step. In Fig. \ref{correlation_time_heisenberg}, 
we show the log-log plot for the relaxation time versus the side length of the cubic simulation cell.  
The dynamical critical exponent is estimated to be around 2.23. This value is comparable to that of the checkerboard MC algorithm (1.96, cf. Ref. \onlinecite{Landau1990}), but cannot compete with the cluster algorithms 
whose relaxation times are almost independent of the size \cite{Holm1993}. However, as we discussed in the introduction, these algorithms are only applicable for a small group of spin models, whereas HMC is generally applicable. 
\subsection{Application to electronic-structure based models}
Next, we employ the magnetic bond-order potential (BOP)\cite{Drautz2011,Matous2011, Ford2015} to demonstrate the application of our algorithm with an auxiliary model, Algorithm II. In algorithm I, we need to run both MC and auxiliary 
spin-dynamics with the target model, which requires too many energy and gradients calls to converge thermal averages for electronic-structure based models. A trajectory length of more than ten is usually needed in auxiliary spin-dynamics in order to decorrelate 
the current and the proposal state. The computational cost of this part can be dramatically reduced with algorithm II in which the auxiliary spin-dynamics is run with a temperature-dependent SCE. 
As the temperature-dependent SCE is orders of magnitude faster than the electronic-structure based models, this gives a significant speed-up.  

The magnetic BOP is one of the simplest electronic-structure based models for magnetic transition metals.
In this model, the potential energy is based on the electronic density of states. It is given as a function of atomic positions and spin orientations,
\begin{equation}
 E(\{\mathbf{r}_i, \mathbf{s}_i \}) = U_{\text{bond}} + U_{\text{trans}} + U_{\text{rep}}+ U_{\text{C}} + U_{\text{X}}+ U_{\text{ext}},
\end{equation}
where $U_{\text{bond}}$ is the bond energy, $U_{\text{trans}}$ the electron transfer energy, $U_{\text{rep}}$ the repulsion energy, $U_{\text{C}}$ the Coulomb energy, $U_{\text{X}}$ the exchange energy, and $U_{\text{ext}}$ the external energy. 
Readers are referred to Ref. \onlinecite{Drautz2011} for a detailed discussion of this model. 

The magnetic BOP gives a robust description of ferromagnetism, and more importantly, the real-material properties such as phase stability, 
elastic constants, and dislocations are described properly \cite{Matous2011, Ford2015, Michael2014}.    
More specifically, we use the 9-moments magnetic BOP of iron by 
Mrovec et al. \cite{Matous2011} and the implementation in the BOPfox code \cite{Hammerschmidt2018} to calculate potential energies for spin configurations in the bcc lattice. The $5\times 5\times5$ simulation cell contains 250 spins, and the atomic positions are fixed.  

\begin{figure}[h!]
\centering
\includegraphics[width=0.9\columnwidth]{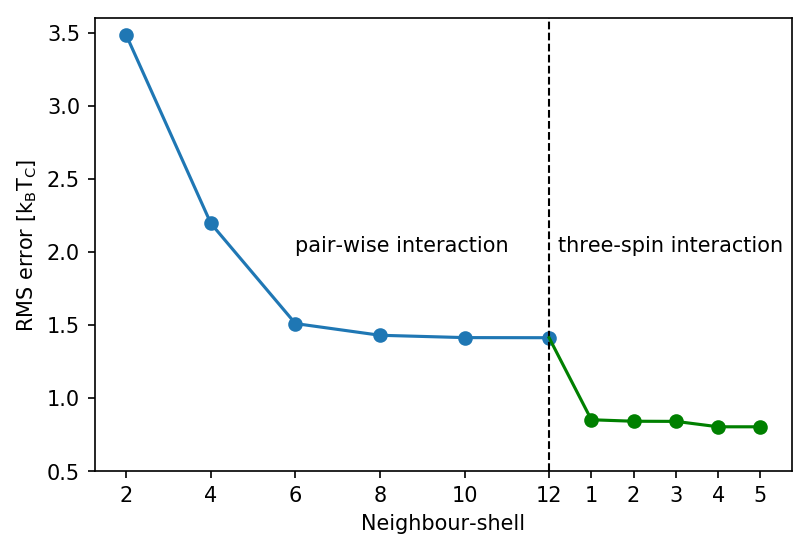}
 \caption{Root-mean-squared (RMS) error of the spin-cluster expansion model plotted as a function of neighbour shells for pair-wise and three-spin interactions. The RMS error per cell is in the unit of $k_\mathrm{B} T_\mathrm{C}$, where $T_\mathrm{C}$ is the 
 experimental Curie temperature of iron (1043~K).  The spin-cluster expansion model is 
 fitted for the potential energy of a magnetic BOP for $5\times5\times5$ bcc lattice of iron.}
 \label{spin_cluster_fitting}
\end{figure}

We employ a temperature-dependent SCE fitted for the BOP of iron as an auxiliary model to run the auxiliary spin-dynamics. As shown in Eq. (\ref{accep_auxiliary}), the energy difference between the electronic-structure based model and the auxiliary model should be as small as possible to guarantee a good 
acceptance probability. A rough estimation can be given for the relationship between the root-mean-squared (RMS) error $\Delta E_{\mathrm{RMS}}$ and the average acceptance probability $\bar{p}^{acc}$,
\begin{equation}
 \bar{p}^{\mathrm{acc}} \approx \frac{1}{2} \left[ 1+\mathrm{exp}\left(-\frac{\Delta E_{\mathrm{RMS}}}{k_{\mathrm{B}}T}\right) \right].
\end{equation}
For example, a RMS error of one $ k_{\mathrm{B}} T$ corresponds to an average acceptance probability of around 0.68, which is a good value for MC acceptance-rejection.  In Fig. \ref{spin_cluster_fitting}, 
we show a typical plot for the RMS error versus neighbour shells of pair-wise and three-spin interactions. 
The pair-wise spin clusters are not sufficient to reduce the root-mean-squared (RMS) error to less than one $k_{\mathrm{B}} T_\mathrm{C}$ and three-spin interactions are 
taken into account in order to further reduce the RMS error to 
0.80~$k_{\mathrm{B}} T_\mathrm{C}$. In practice, we include pair-wise interactions up to the sixth nearest-neighbour shell and 
the first-nearest-neighbour three-spin interactions in our temperature-dependent SCE. 
The two-spin and three-spin interactions are sufficient to converge the SCE for the ideal lattice with fixed atomic positions that is used in this work.
Breaking the geometric degeneracy by, e.g., vibrations or defects, will introduce distance- and environment-dependent interaction terms which rule the convergence of the SCE much more difficult or even intractable.

\begin{figure}[h!]
\centering
\includegraphics[width=0.9\columnwidth]{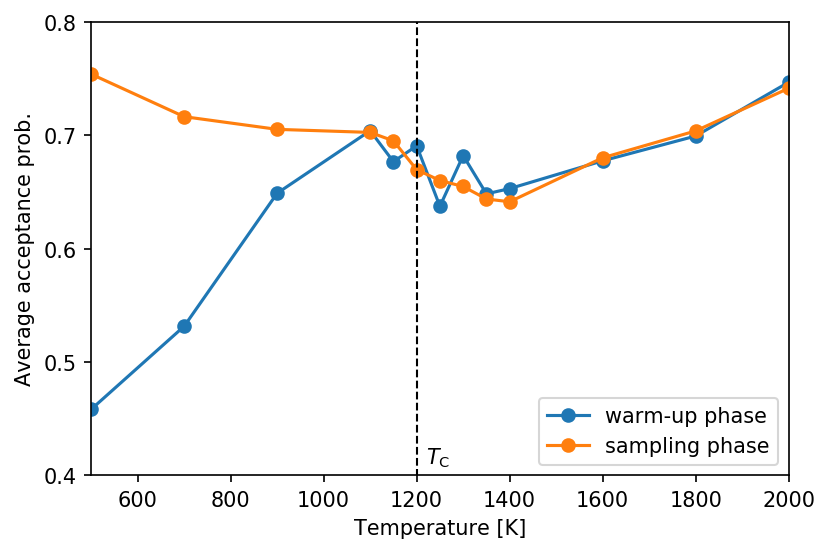}
 \caption{Average acceptance probability in the warm-up and the sampling phases. The temperature-independent spin cluster expansion is used in the warm-up phase while the temperature-dependent one used in the sampling phase. The magnetic BOP of iron is the target model 
 in this test, with $5\times5\times5$ bcc lattice of 250 spins.}
 \label{average_acceptance_prob}
\end{figure}

As discussed in Appendix \ref{implementation_detail}, we employ the temperature-independent and temperature-dependent 
SCE in the warm-up phase and sampling phase, respectively. In Fig. \ref{average_acceptance_prob}, 
\begin{figure}[h!]
\centering
\includegraphics[width=0.9\columnwidth]{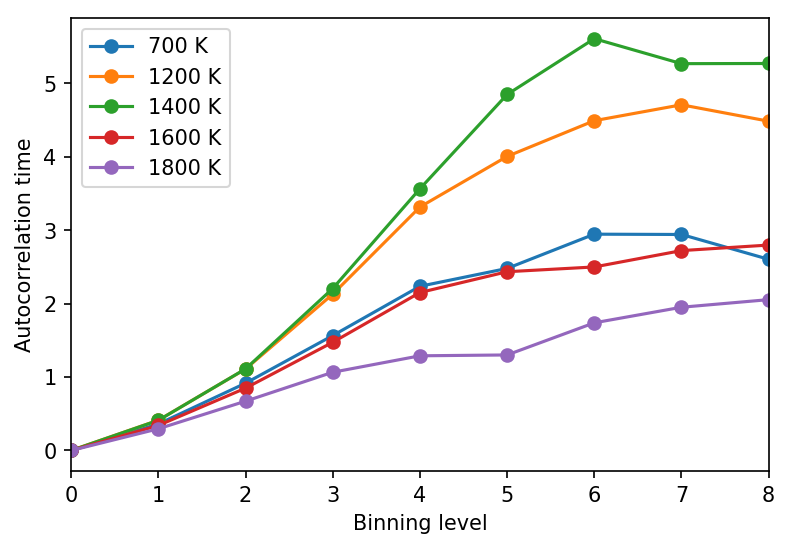}
 \caption{Binning analysis for potential energies of $5\times5\times5$ bcc lattice at different temperatures.}
 \label{BOP_autocorrelation_time}
\end{figure}
we show the average acceptance probability at different temperatures in both the warm-up and sampling phase. 
It is higher in the sampling phase than in the warm-up phase as the temperature-dependent SCE reproduces the energy 
landscape of the BOP better, as discussed in Sec. \ref{methodology_section}.

\begin{figure}[h!]
\centering
\includegraphics[width=0.9\columnwidth]{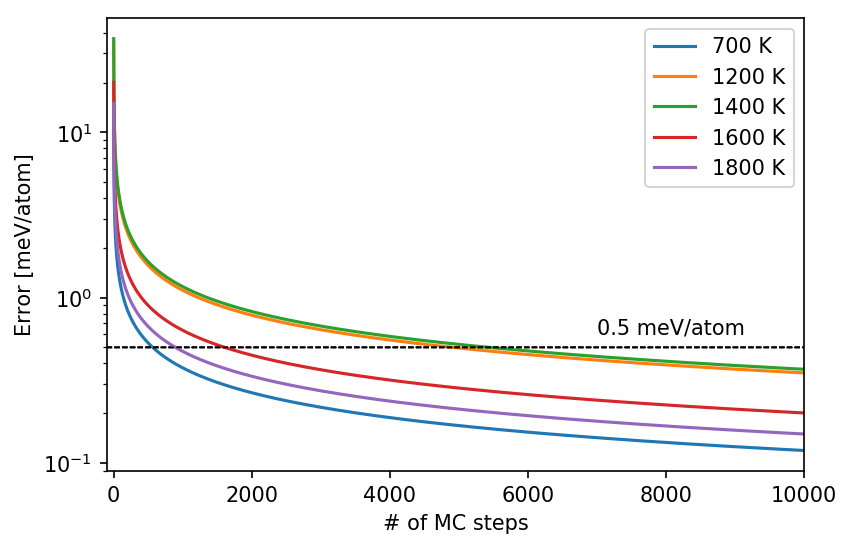}
 \caption{Running errors in the estimation of thermal averages of potential energies at different temperatures.}
 \label{BOP_running_error}
\end{figure}

We use 10,000 MC steps to evaluate thermal averages, and perform a binning analysis \cite{Ambegaokar2009} for the potential energy to check the convergence. As shown in Fig. \ref{BOP_autocorrelation_time}, 
the autocorrelation times at all temperatures reach a plateau, which indicates that the calculation is fully converged. The autocorrelation times range from two to six depending on the temperature. The running error can be estimated according to\cite{Ambegaokar2009}
\begin{equation}
 \Delta_O = \sqrt{\frac{\mathrm{Var}O}{N} (1+2\tau_O)},
\end{equation}
where $O$ is the observable, $\mathrm{Var}O$ is its variance, $\tau_O$ is its autocorrelation time, and $N$ is the number of MC steps. Based on this, we plot in Fig. \ref{BOP_running_error} the running errors in the estimation of thermal averages of potential energies at different temperatures. 
Clearly, more MC steps are required to reach convergence at temperatures closer to the critical point (1200~K in this case, as shown later) as the variance is larger.  For our algorithm, an error of 0.5~meV/atom can be reached within 5000 MC steps. This excellent efficiency is due to 
the short autocorrelation time, as already discussed.  

\begin{figure}[h!]
\centering
\includegraphics[width=0.9\columnwidth]{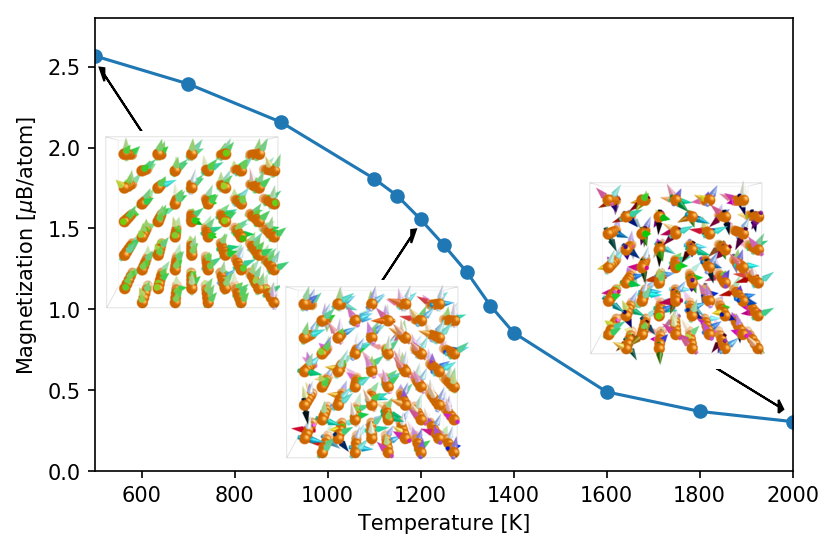}
 \caption{Magnetization as a function of temperature for $5\times5\times5$ bcc lattice of iron obtained with a magnetic bond-order potential. The images are snapshots of spin configurations at 500~K, 1200~K, and 2000~K, respectively, 
 which are generated with the code V\_sim \cite{Vsim}. The spin directions are indicated by the arrows and the coloration.
 We see the collapse of the long- and short-range magnetic orders with increasing temperatures.  }
 \label{BOP_magnetization}
\end{figure}

\begin{figure}[h!]
\centering
\includegraphics[width=0.9\columnwidth]{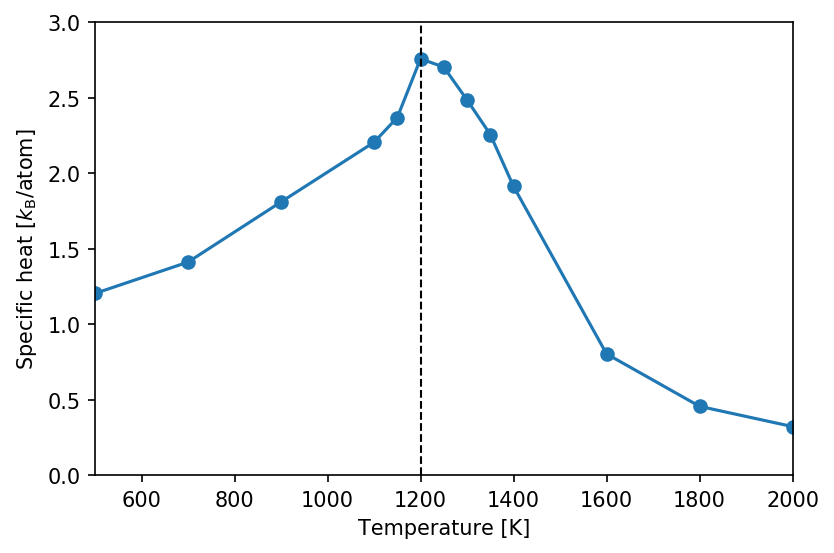}
 \caption{Magnetic contribution to the specific heat as a function of temperatures for $5\times5\times5$ bcc lattice of iron for magnetic bond-order potentials.}
 \label{BOP_specific_heat}
\end{figure}

The magnetization and the magnetic contribution to the specific heat of bcc iron for magnetic BOP are plotted as a function of temperature
in Fig. \ref{BOP_magnetization} and Fig. \ref{BOP_specific_heat}, respectively.
The magnetization curve we obtain here has similar features to those of the classical Heisenberg model \cite{Ma2012, Fritz2011}, which indicates that 
the exchange parameters are not influenced considerably by spin fluctuations in magnetic iron. We notice that there is a recent paper by Ruban and Peil \cite{Ruban2018} who found that the exchange parameters are significantly influenced by atomic vibrations in magnetic iron. 
This effect is not considered here as we fix atomic positions in our simulation. There is residual magnetization up to 2000~K due to the finite-size effect. Based on the magnetic contribution to the specific heat (see Fig. \ref{BOP_specific_heat}), we estimate the Curie temperature to be around 1200~K, 
which is close to the experimental value (1043~K). The difference in $T_\mathrm{C}$ is 
attributed to the neglect of magnon-phonon coupling \cite{Ruban2018} and limitations of the parametrization of the magnetic BOP used here. 
To our knowledge, this is the first result for the direct and high-fidelity simulations of the magnetic phase transition of bcc iron with an electronic-structure based model, in contrast to related works in literature 
that are based on parametrized model Hamiltonians \cite{Kormann2011, Ma2012, Ruban2018}.

\section{Conclusions}
We have developed a Hamiltonian Monte Carlo (HMC) framework in order to efficiently sample the spin space for electronic-structure based models of magnetic materials. 
From auxiliary spin-dynamics we derive a HMC algorithm for spin systems, which is denoted as Algorithm I in this paper. The hyper-parameters (the time step and the trajectory length) are tuned with automatic tuning procedures without user intervention. Our tests with the classical Heisenberg model
show that with the automatic tunning procedures of hyper-parameters this algorithm has a fast warm-up efficiency. The dynamical critical exponent is estimated to be $2.23\pm 0.02$, close to that of the checkerboard MC algorithm. 
While the latter is only applicable for a limited group of spin models, our algorithm is of general applicability.     
The utilization of the temperature-dependent spin-cluster expansion as an auxiliary model to run auxiliary spin-dynamics further accelerates the exploration of the spin space, and based on this we develop Algorithm II in this paper.   
Our application employing the magnetic bond-order potentials demonstrates the efficiency of our sampler. For a 250-atom supercell of bcc iron, the autocorrelation time of the potential energy is less than six MC steps, and the thermal average can be converged with an error of 
$\pm0.5$~meV/atom within 6000 steps in the whole temperature range. 
We compute the magnetization curve, the magnetic contribution to the specific heat, and the Curie temperature of bcc iron as predicted by the BOP model with high fidelity.  
We conclude that this work paves the way towards atomistic simulations of magnetic materials 
with complex spin interactions, and look forward to seeing applications of our method for more complex models, such as non-collinear magnetic density-functional theory.

\section*{ACKNOWLEDGEMENT}
NW acknowledges a PhD fellowship from the International Max Planck Research School for Surface and Interface Engineering (IMPRS-SurMat). The authors thank Martin Staadt and Tilmann Hickel for fruitful discussions.

\section*{APPENDIX}
\appendix

\section{U-turn termination criterion}\label{U_turn_termination}
The U-turn termination criterion\cite{Hoffman2014} is an empirical estimate for the optimal length of Hamiltonian dynamics per MC step. The basic idea is to maximize the squared distance between the initial and final states. 
Its implementation in this work is sightly different from the one in the original paper as we do not use standard molecular dynamics. 

We first define the half squared distance for spin systems,  
\begin{equation}\label{half_squared_distance}
 \Delta(t) = \frac{1}{2} \sum_i \left[ \mathbf{s}_i (t) -\mathbf{s}_i (0) \right] ^T \cdot \left[ \mathbf{s}_i (t) -\mathbf{s}_i (0) \right].
\end{equation}
The U-turn termination criterion is then derived according to,
\begin{equation}
\begin{aligned}
 & \frac{d\Delta}{dt} < 0,
 \end{aligned}
\end{equation}
and given as, 
\begin{equation}
\begin{aligned}
 & \sum_i \left[ \mathbf{s}_i(t) - \mathbf{s}_i(0) \right]^T \cdot [  \bm{\omega}_i(t) \times \mathbf{s}_i(t) ] < 0,
\end{aligned}
\end{equation}
where the equations of motion, Eq. (\ref{equations_of_motion}), are employed. 

\section{Implementation details} \label{implementation_detail}
We first fit a temperature-independent spin-cluster expansion as a start-up auxiliary model. This involves 1000 spin configurations generated with the classical Heisenberg model at the critical point and 
the corresponding potential energies for electronic-structure based models (magnetic bond-order potential in our application). 
The warm-up phase is split into two stages. In the first stage, the system is thermalized purely with the start-up auxiliary model and the length is set to 1000 MC steps. 
The time-step and the trajectory length are automatically tuned with the methods in Section \ref{methodology_section}.
In the second stage, the start-up model is used to run auxiliary spin-dynamics while the Metropolis-Hastings acceptance-rejection is performed for the target electronic structure based models. 
The time-step is further tuned while the trajectory length is fixed to the mean value in the first stage. In our implementation for the magnetic bond-order potential, a typical time step is around 0.3 fs, and a typical trajectory length is around 18.
The length of the second stage is set to be 500 MC steps. Once the warm-up phase is finished, we collect all the spin configurations and the potential energies in the second stage to fit a temperature-dependent SCE with least-squares fitting, which is then used to run auxiliary spin-dynamics 
in the sampling phase. Both the time-step and the trajectory length are fixed in this phase. We use 10,000 MC steps to evaluate thermal averages, which are sufficient to guarantee convergence due to the small autocorrelation time in our algorithm. 

\bibliography{paper_hybridMC}

\begin{thebibliography}{61}
\expandafter\ifx\csname natexlab\endcsname\relax\def\natexlab#1{#1}\fi
\expandafter\ifx\csname bibnamefont\endcsname\relax
  \def\bibnamefont#1{#1}\fi
\expandafter\ifx\csname bibfnamefont\endcsname\relax
  \def\bibfnamefont#1{#1}\fi
\expandafter\ifx\csname citenamefont\endcsname\relax
  \def\citenamefont#1{#1}\fi
\expandafter\ifx\csname url\endcsname\relax
  \def\url#1{\texttt{#1}}\fi
\expandafter\ifx\csname urlprefix\endcsname\relax\def\urlprefix{URL }\fi
\providecommand{\bibinfo}[2]{#2}
\providecommand{\eprint}[2][]{\url{#2}}

\bibitem[{\citenamefont{Pecharsky and Jr}(1999)}]{Vitalij1999}
\bibinfo{author}{\bibfnamefont{V.~K.} \bibnamefont{Pecharsky}}
  \bibnamefont{and} \bibinfo{author}{\bibfnamefont{K.~A.~G.} \bibnamefont{Jr}},
  \bibinfo{journal}{Journal of Magnetism and Magnetic Materials}
  \textbf{\bibinfo{volume}{200}}, \bibinfo{pages}{44 } (\bibinfo{year}{1999}).

\bibitem[{\citenamefont{Hickel et~al.}(2012)\citenamefont{Hickel, Grabowski,
  K{\"o}rmann, and Neugebauer}}]{Hickel2012}
\bibinfo{author}{\bibfnamefont{T.}~\bibnamefont{Hickel}},
  \bibinfo{author}{\bibfnamefont{B.}~\bibnamefont{Grabowski}},
  \bibinfo{author}{\bibfnamefont{F.}~\bibnamefont{K{\"o}rmann}},
  \bibnamefont{and}
  \bibinfo{author}{\bibfnamefont{J.}~\bibnamefont{Neugebauer}},
  \bibinfo{journal}{Journal of Physics: Condensed Matter}
  \textbf{\bibinfo{volume}{24}}, \bibinfo{pages}{053202}
  (\bibinfo{year}{2012}).

\bibitem[{\citenamefont{Lu et~al.}(2007)\citenamefont{Lu, Salabas, and
  Sch\"uth}}]{Lu2007}
\bibinfo{author}{\bibfnamefont{A.-H.} \bibnamefont{Lu}},
  \bibinfo{author}{\bibfnamefont{E.~L.} \bibnamefont{Salabas}},
  \bibnamefont{and} \bibinfo{author}{\bibfnamefont{F.}~\bibnamefont{Sch\"uth}},
  \bibinfo{journal}{Angewandte Chemie International Edition}
  \textbf{\bibinfo{volume}{46}}, \bibinfo{pages}{1222} (\bibinfo{year}{2007}).

\bibitem[{\citenamefont{Dietermann et~al.}(2012)\citenamefont{Dietermann,
  Sandratskii, and F{\"a}hnle}}]{Dietermann2012}
\bibinfo{author}{\bibfnamefont{F.}~\bibnamefont{Dietermann}},
  \bibinfo{author}{\bibfnamefont{L.}~\bibnamefont{Sandratskii}},
  \bibnamefont{and}
  \bibinfo{author}{\bibfnamefont{M.}~\bibnamefont{F{\"a}hnle}},
  \bibinfo{journal}{Journal of Magnetism and Magnetic Materials}
  \textbf{\bibinfo{volume}{324}}, \bibinfo{pages}{2693 }
  (\bibinfo{year}{2012}), ISSN \bibinfo{issn}{0304-8853}.

\bibitem[{\citenamefont{Ruban et~al.}(2007)\citenamefont{Ruban, Khmelevskyi,
  Mohn, and Johansson}}]{Ruban2007}
\bibinfo{author}{\bibfnamefont{A.~V.} \bibnamefont{Ruban}},
  \bibinfo{author}{\bibfnamefont{S.}~\bibnamefont{Khmelevskyi}},
  \bibinfo{author}{\bibfnamefont{P.}~\bibnamefont{Mohn}}, \bibnamefont{and}
  \bibinfo{author}{\bibfnamefont{B.}~\bibnamefont{Johansson}},
  \bibinfo{journal}{Phys. Rev. B} \textbf{\bibinfo{volume}{75}},
  \bibinfo{pages}{054402} (\bibinfo{year}{2007}).

\bibitem[{\citenamefont{Ma and Dudarev}(2012{\natexlab{a}})}]{Dudarev2012}
\bibinfo{author}{\bibfnamefont{P.-W.} \bibnamefont{Ma}} \bibnamefont{and}
  \bibinfo{author}{\bibfnamefont{S.~L.} \bibnamefont{Dudarev}},
  \bibinfo{journal}{Phys. Rev. B} \textbf{\bibinfo{volume}{86}},
  \bibinfo{pages}{054416} (\bibinfo{year}{2012}{\natexlab{a}}).

\bibitem[{\citenamefont{Ma et~al.}(2008)\citenamefont{Ma, Woo, and
  Dudarev}}]{Ma2008}
\bibinfo{author}{\bibfnamefont{P.-W.} \bibnamefont{Ma}},
  \bibinfo{author}{\bibfnamefont{C.~H.} \bibnamefont{Woo}}, \bibnamefont{and}
  \bibinfo{author}{\bibfnamefont{S.~L.} \bibnamefont{Dudarev}},
  \bibinfo{journal}{Phys. Rev. B} \textbf{\bibinfo{volume}{78}},
  \bibinfo{pages}{024434} (\bibinfo{year}{2008}).

\bibitem[{\citenamefont{Moriya}(1985)}]{Moriya1985}
\bibinfo{author}{\bibfnamefont{T.}~\bibnamefont{Moriya}},
  \emph{\bibinfo{title}{Spin fluctuations in itinerant electron magnetism}},
  Springer series in solid-state sciences
  (\bibinfo{publisher}{Springer-Verlag}, \bibinfo{year}{1985}), ISBN
  \bibinfo{isbn}{9783540154228}.

\bibitem[{\citenamefont{Kvashnin et~al.}(2016)\citenamefont{Kvashnin, Cardias,
  Szilva, Di~Marco, Katsnelson, Lichtenstein, Nordstr\"om, Klautau, and
  Eriksson}}]{Kvashnin2016}
\bibinfo{author}{\bibfnamefont{Y.~O.} \bibnamefont{Kvashnin}},
  \bibinfo{author}{\bibfnamefont{R.}~\bibnamefont{Cardias}},
  \bibinfo{author}{\bibfnamefont{A.}~\bibnamefont{Szilva}},
  \bibinfo{author}{\bibfnamefont{I.}~\bibnamefont{Di~Marco}},
  \bibinfo{author}{\bibfnamefont{M.~I.} \bibnamefont{Katsnelson}},
  \bibinfo{author}{\bibfnamefont{A.~I.} \bibnamefont{Lichtenstein}},
  \bibinfo{author}{\bibfnamefont{L.}~\bibnamefont{Nordstr\"om}},
  \bibinfo{author}{\bibfnamefont{A.~B.} \bibnamefont{Klautau}},
  \bibnamefont{and} \bibinfo{author}{\bibfnamefont{O.}~\bibnamefont{Eriksson}},
  \bibinfo{journal}{Phys. Rev. Lett.} \textbf{\bibinfo{volume}{116}},
  \bibinfo{pages}{217202} (\bibinfo{year}{2016}).

\bibitem[{\citenamefont{K{\"o}rmann et~al.}(2014)\citenamefont{K{\"o}rmann,
  Grabowski, Dutta, Hickel, Mauger, Fultz, and Neugebauer}}]{Fritz2014}
\bibinfo{author}{\bibfnamefont{F.}~\bibnamefont{K{\"o}rmann}},
  \bibinfo{author}{\bibfnamefont{B.}~\bibnamefont{Grabowski}},
  \bibinfo{author}{\bibfnamefont{B.}~\bibnamefont{Dutta}},
  \bibinfo{author}{\bibfnamefont{T.}~\bibnamefont{Hickel}},
  \bibinfo{author}{\bibfnamefont{L.}~\bibnamefont{Mauger}},
  \bibinfo{author}{\bibfnamefont{B.}~\bibnamefont{Fultz}}, \bibnamefont{and}
  \bibinfo{author}{\bibfnamefont{J.}~\bibnamefont{Neugebauer}},
  \bibinfo{journal}{Phys. Rev. Lett.} \textbf{\bibinfo{volume}{113}},
  \bibinfo{pages}{165503} (\bibinfo{year}{2014}).

\bibitem[{\citenamefont{Kubler et~al.}(1988)\citenamefont{Kubler, Hock, Sticht,
  and Williams}}]{Kubler1988}
\bibinfo{author}{\bibfnamefont{J.}~\bibnamefont{Kubler}},
  \bibinfo{author}{\bibfnamefont{K.~H.} \bibnamefont{Hock}},
  \bibinfo{author}{\bibfnamefont{J.}~\bibnamefont{Sticht}}, \bibnamefont{and}
  \bibinfo{author}{\bibfnamefont{A.~R.} \bibnamefont{Williams}},
  \bibinfo{journal}{Journal of Physics F: Metal Physics}
  \textbf{\bibinfo{volume}{18}}, \bibinfo{pages}{469} (\bibinfo{year}{1988}).

\bibitem[{\citenamefont{Stocks et~al.}(1998)\citenamefont{Stocks, Ujfalussy,
  Wang, Nicholson, Shelton, Wang, Canning, and Gy{\"o}rffy}}]{Stocks1998}
\bibinfo{author}{\bibfnamefont{G.~M.} \bibnamefont{Stocks}},
  \bibinfo{author}{\bibfnamefont{B.}~\bibnamefont{Ujfalussy}},
  \bibinfo{author}{\bibfnamefont{X.}~\bibnamefont{Wang}},
  \bibinfo{author}{\bibfnamefont{D.~M.~C.} \bibnamefont{Nicholson}},
  \bibinfo{author}{\bibfnamefont{W.~A.} \bibnamefont{Shelton}},
  \bibinfo{author}{\bibfnamefont{Y.}~\bibnamefont{Wang}},
  \bibinfo{author}{\bibfnamefont{A.}~\bibnamefont{Canning}}, \bibnamefont{and}
  \bibinfo{author}{\bibfnamefont{B.~L.} \bibnamefont{Gy{\"o}rffy}},
  \bibinfo{journal}{Philosophical Magazine B} \textbf{\bibinfo{volume}{78}},
  \bibinfo{pages}{665} (\bibinfo{year}{1998}).

\bibitem[{\citenamefont{Hobbs et~al.}(2000)\citenamefont{Hobbs, Kresse, and
  Hafner}}]{Hobbs2000}
\bibinfo{author}{\bibfnamefont{D.}~\bibnamefont{Hobbs}},
  \bibinfo{author}{\bibfnamefont{G.}~\bibnamefont{Kresse}}, \bibnamefont{and}
  \bibinfo{author}{\bibfnamefont{J.}~\bibnamefont{Hafner}},
  \bibinfo{journal}{Phys. Rev. B} \textbf{\bibinfo{volume}{62}},
  \bibinfo{pages}{11556} (\bibinfo{year}{2000}).

\bibitem[{\citenamefont{Kurz et~al.}(2004)\citenamefont{Kurz, F\"orster,
  Nordstr\"om, Bihlmayer, and Bl\"ugel}}]{Blugel2004}
\bibinfo{author}{\bibfnamefont{P.}~\bibnamefont{Kurz}},
  \bibinfo{author}{\bibfnamefont{F.}~\bibnamefont{F\"orster}},
  \bibinfo{author}{\bibfnamefont{L.}~\bibnamefont{Nordstr\"om}},
  \bibinfo{author}{\bibfnamefont{G.}~\bibnamefont{Bihlmayer}},
  \bibnamefont{and} \bibinfo{author}{\bibfnamefont{S.}~\bibnamefont{Bl\"ugel}},
  \bibinfo{journal}{Phys. Rev. B} \textbf{\bibinfo{volume}{69}},
  \bibinfo{pages}{024415} (\bibinfo{year}{2004}).

\bibitem[{\citenamefont{Peralta et~al.}(2007)\citenamefont{Peralta, Scuseria,
  and Frisch}}]{Peralta2007}
\bibinfo{author}{\bibfnamefont{J.~E.} \bibnamefont{Peralta}},
  \bibinfo{author}{\bibfnamefont{G.~E.} \bibnamefont{Scuseria}},
  \bibnamefont{and} \bibinfo{author}{\bibfnamefont{M.~J.}
  \bibnamefont{Frisch}}, \bibinfo{journal}{Phys. Rev. B}
  \textbf{\bibinfo{volume}{75}}, \bibinfo{pages}{125119}
  (\bibinfo{year}{2007}).

\bibitem[{\citenamefont{Ma and Dudarev}(2015)}]{Ma2015}
\bibinfo{author}{\bibfnamefont{P.-W.} \bibnamefont{Ma}} \bibnamefont{and}
  \bibinfo{author}{\bibfnamefont{S.~L.} \bibnamefont{Dudarev}},
  \bibinfo{journal}{Phys. Rev. B} \textbf{\bibinfo{volume}{91}},
  \bibinfo{pages}{054420} (\bibinfo{year}{2015}).

\bibitem[{\citenamefont{Mukherjee and Cohen}(2001)}]{Mukherjee2001}
\bibinfo{author}{\bibfnamefont{S.}~\bibnamefont{Mukherjee}} \bibnamefont{and}
  \bibinfo{author}{\bibfnamefont{R.}~\bibnamefont{Cohen}},
  \bibinfo{journal}{Journal of Computer-Aided Materials Design}
  \textbf{\bibinfo{volume}{8}}, \bibinfo{pages}{107} (\bibinfo{year}{2001}).

\bibitem[{\citenamefont{Barreteau et~al.}(2016)\citenamefont{Barreteau,
  Spanjaard, and Desjonqu\'eres}}]{Barreteau2015}
\bibinfo{author}{\bibfnamefont{C.}~\bibnamefont{Barreteau}},
  \bibinfo{author}{\bibfnamefont{D.}~\bibnamefont{Spanjaard}},
  \bibnamefont{and} \bibinfo{author}{\bibfnamefont{M.-C.}
  \bibnamefont{Desjonqu\'eres}}, \bibinfo{journal}{Comptes Rendus Physique}
  \textbf{\bibinfo{volume}{17}}, \bibinfo{pages}{406 } (\bibinfo{year}{2016}).

\bibitem[{\citenamefont{Drautz and Pettifor}(2011)}]{Drautz2011}
\bibinfo{author}{\bibfnamefont{R.}~\bibnamefont{Drautz}} \bibnamefont{and}
  \bibinfo{author}{\bibfnamefont{D.~G.} \bibnamefont{Pettifor}},
  \bibinfo{journal}{Phys. Rev. B} \textbf{\bibinfo{volume}{84}},
  \bibinfo{pages}{214114} (\bibinfo{year}{2011}).

\bibitem[{\citenamefont{Ford et~al.}(2014)\citenamefont{Ford, Drautz,
  Hammerschmidt, and Pettifor}}]{Michael2014}
\bibinfo{author}{\bibfnamefont{M.~E.} \bibnamefont{Ford}},
  \bibinfo{author}{\bibfnamefont{R.}~\bibnamefont{Drautz}},
  \bibinfo{author}{\bibfnamefont{T.}~\bibnamefont{Hammerschmidt}},
  \bibnamefont{and} \bibinfo{author}{\bibfnamefont{D.~G.}
  \bibnamefont{Pettifor}}, \bibinfo{journal}{Modelling and Simulation in
  Materials Science and Engineering} \textbf{\bibinfo{volume}{22}},
  \bibinfo{pages}{034005} (\bibinfo{year}{2014}).

\bibitem[{\citenamefont{Ford et~al.}(2015)\citenamefont{Ford, Pettifor, and
  Drautz}}]{Ford2015}
\bibinfo{author}{\bibfnamefont{M.~E.} \bibnamefont{Ford}},
  \bibinfo{author}{\bibfnamefont{D.~G.} \bibnamefont{Pettifor}},
  \bibnamefont{and} \bibinfo{author}{\bibfnamefont{R.}~\bibnamefont{Drautz}},
  \bibinfo{journal}{Journal of Physics: Condensed Matter}
  \textbf{\bibinfo{volume}{27}}, \bibinfo{pages}{086002}
  (\bibinfo{year}{2015}).

\bibitem[{\citenamefont{Drautz et~al.}(2015)\citenamefont{Drautz,
  Hammerschmidt, \v{C}\'{a}k, and Pettifor}}]{Drautz2015}
\bibinfo{author}{\bibfnamefont{R.}~\bibnamefont{Drautz}},
  \bibinfo{author}{\bibfnamefont{T.}~\bibnamefont{Hammerschmidt}},
  \bibinfo{author}{\bibfnamefont{M.}~\bibnamefont{\v{C}\'{a}k}},
  \bibnamefont{and} \bibinfo{author}{\bibfnamefont{D.~G.}
  \bibnamefont{Pettifor}}, \bibinfo{journal}{Modelling and Simulation in
  Materials Science and Engineering} \textbf{\bibinfo{volume}{23}},
  \bibinfo{pages}{074004} (\bibinfo{year}{2015}).

\bibitem[{\citenamefont{Peczak and Landau}(1990)}]{Landau1990}
\bibinfo{author}{\bibfnamefont{P.}~\bibnamefont{Peczak}} \bibnamefont{and}
  \bibinfo{author}{\bibfnamefont{D.~P.} \bibnamefont{Landau}},
  \bibinfo{journal}{Journal of Applied Physics} \textbf{\bibinfo{volume}{67}},
  \bibinfo{pages}{5427} (\bibinfo{year}{1990}).

\bibitem[{\citenamefont{Swendsen and Wang}(1987)}]{Swendsen1987}
\bibinfo{author}{\bibfnamefont{R.~H.} \bibnamefont{Swendsen}} \bibnamefont{and}
  \bibinfo{author}{\bibfnamefont{J.-S.} \bibnamefont{Wang}},
  \bibinfo{journal}{Phys. Rev. Lett.} \textbf{\bibinfo{volume}{58}},
  \bibinfo{pages}{86} (\bibinfo{year}{1987}).

\bibitem[{\citenamefont{Wolff}(1989)}]{Wolff1989}
\bibinfo{author}{\bibfnamefont{U.}~\bibnamefont{Wolff}},
  \bibinfo{journal}{Phys. Rev. Lett.} \textbf{\bibinfo{volume}{62}},
  \bibinfo{pages}{361} (\bibinfo{year}{1989}).

\bibitem[{\citenamefont{Brown and Woch}(1987)}]{Brown1987}
\bibinfo{author}{\bibfnamefont{F.~R.} \bibnamefont{Brown}} \bibnamefont{and}
  \bibinfo{author}{\bibfnamefont{T.~J.} \bibnamefont{Woch}},
  \bibinfo{journal}{Phys. Rev. Lett.} \textbf{\bibinfo{volume}{58}},
  \bibinfo{pages}{2394} (\bibinfo{year}{1987}).

\bibitem[{\citenamefont{Creutz}(1987)}]{Creutz1987}
\bibinfo{author}{\bibfnamefont{M.}~\bibnamefont{Creutz}},
  \bibinfo{journal}{Phys. Rev. D} \textbf{\bibinfo{volume}{36}},
  \bibinfo{pages}{515} (\bibinfo{year}{1987}).

\bibitem[{\citenamefont{Landau and Binder}(2005)}]{Landau2005}
\bibinfo{author}{\bibfnamefont{D.}~\bibnamefont{Landau}} \bibnamefont{and}
  \bibinfo{author}{\bibfnamefont{K.}~\bibnamefont{Binder}},
  \emph{\bibinfo{title}{A Guide to Monte Carlo Simulations in Statistical
  Physics}} (\bibinfo{publisher}{Cambridge University Press},
  \bibinfo{address}{New York, NY, USA}, \bibinfo{year}{2005}), ISBN
  \bibinfo{isbn}{0521842387}.

\bibitem[{\citenamefont{Ma and Dudarev}(2011)}]{Dudarev2011}
\bibinfo{author}{\bibfnamefont{P.-W.} \bibnamefont{Ma}} \bibnamefont{and}
  \bibinfo{author}{\bibfnamefont{S.~L.} \bibnamefont{Dudarev}},
  \bibinfo{journal}{Phys. Rev. B} \textbf{\bibinfo{volume}{83}},
  \bibinfo{pages}{134418} (\bibinfo{year}{2011}).

\bibitem[{\citenamefont{Tranchida et~al.}(2018)\citenamefont{Tranchida,
  Plimpton, Thibaudeau, and Thompson}}]{Tranchida2018}
\bibinfo{author}{\bibfnamefont{J.}~\bibnamefont{Tranchida}},
  \bibinfo{author}{\bibfnamefont{S.}~\bibnamefont{Plimpton}},
  \bibinfo{author}{\bibfnamefont{P.}~\bibnamefont{Thibaudeau}},
  \bibnamefont{and} \bibinfo{author}{\bibfnamefont{A.}~\bibnamefont{Thompson}},
  \bibinfo{journal}{Journal of Computational Physics}
  \textbf{\bibinfo{volume}{372}}, \bibinfo{pages}{406 } (\bibinfo{year}{2018}).

\bibitem[{\citenamefont{Wang and Landau}(2001)}]{Wang2001}
\bibinfo{author}{\bibfnamefont{F.}~\bibnamefont{Wang}} \bibnamefont{and}
  \bibinfo{author}{\bibfnamefont{D.~P.} \bibnamefont{Landau}},
  \bibinfo{journal}{Phys. Rev. Lett.} \textbf{\bibinfo{volume}{86}},
  \bibinfo{pages}{2050} (\bibinfo{year}{2001}).

\bibitem[{\citenamefont{Eisenbach et~al.}(2011)\citenamefont{Eisenbach,
  Nicholson, Rusanu, and Brown}}]{Eisenbach2011}
\bibinfo{author}{\bibfnamefont{M.}~\bibnamefont{Eisenbach}},
  \bibinfo{author}{\bibfnamefont{D.~M.} \bibnamefont{Nicholson}},
  \bibinfo{author}{\bibfnamefont{A.}~\bibnamefont{Rusanu}}, \bibnamefont{and}
  \bibinfo{author}{\bibfnamefont{G.}~\bibnamefont{Brown}},
  \bibinfo{journal}{Journal of Applied Physics} \textbf{\bibinfo{volume}{109}},
  \bibinfo{pages}{07E138} (\bibinfo{year}{2011}).

\bibitem[{\citenamefont{Neal}(2011)}]{Neal2011}
\bibinfo{author}{\bibfnamefont{R.~M.} \bibnamefont{Neal}}, in
  \emph{\bibinfo{booktitle}{Handbook of Markov chain Monte Carlo}}, edited by
  \bibinfo{editor}{\bibfnamefont{S.}~\bibnamefont{Brooks}},
  \bibinfo{editor}{\bibfnamefont{A.}~\bibnamefont{Gelman}},
  \bibinfo{editor}{\bibfnamefont{G.~L.} \bibnamefont{Jones}}, \bibnamefont{and}
  \bibinfo{editor}{\bibfnamefont{X.-L.} \bibnamefont{Meng}}
  (\bibinfo{publisher}{Chapman \& Hall/CRC}, \bibinfo{year}{2011}),
  chap.~\bibinfo{chapter}{5}, pp. \bibinfo{pages}{113--162}.

\bibitem[{\citenamefont{Mark and Ben}(2011)}]{Mark2011}
\bibinfo{author}{\bibfnamefont{G.}~\bibnamefont{Mark}} \bibnamefont{and}
  \bibinfo{author}{\bibfnamefont{C.}~\bibnamefont{Ben}},
  \bibinfo{journal}{Journal of the Royal Statistical Society: Series B
  (Statistical Methodology)} \textbf{\bibinfo{volume}{73}},
  \bibinfo{pages}{123} (\bibinfo{year}{2011}).

\bibitem[{\citenamefont{Beskos et~al.}(2013)\citenamefont{Beskos, Pillai,
  Roberts, Sanz-Serna, and Stuart}}]{beskos2013}
\bibinfo{author}{\bibfnamefont{A.}~\bibnamefont{Beskos}},
  \bibinfo{author}{\bibfnamefont{N.}~\bibnamefont{Pillai}},
  \bibinfo{author}{\bibfnamefont{G.}~\bibnamefont{Roberts}},
  \bibinfo{author}{\bibfnamefont{J.-M.} \bibnamefont{Sanz-Serna}},
  \bibnamefont{and} \bibinfo{author}{\bibfnamefont{A.}~\bibnamefont{Stuart}},
  \bibinfo{journal}{Bernoulli} \textbf{\bibinfo{volume}{19}},
  \bibinfo{pages}{1501} (\bibinfo{year}{2013}).

\bibitem[{\citenamefont{{Betancourt}}(2017)}]{Betancourt2017_2}
\bibinfo{author}{\bibfnamefont{M.}~\bibnamefont{{Betancourt}}},
  \bibinfo{journal}{ArXiv e-prints}  (\bibinfo{year}{2017}),
  \eprint{1701.02434}.

\bibitem[{\citenamefont{Betancourt et~al.}(2017)\citenamefont{Betancourt,
  Byrne, Livingstone, and Girolami}}]{Betancourt2017_1}
\bibinfo{author}{\bibfnamefont{M.}~\bibnamefont{Betancourt}},
  \bibinfo{author}{\bibfnamefont{S.}~\bibnamefont{Byrne}},
  \bibinfo{author}{\bibfnamefont{S.}~\bibnamefont{Livingstone}},
  \bibnamefont{and} \bibinfo{author}{\bibfnamefont{M.}~\bibnamefont{Girolami}},
  \bibinfo{journal}{Bernoulli} \textbf{\bibinfo{volume}{23}},
  \bibinfo{pages}{2257} (\bibinfo{year}{2017}).

\bibitem[{\citenamefont{Wang et~al.}(2013)\citenamefont{Wang, Mohamed, and
  De~Freitas}}]{Wang2013}
\bibinfo{author}{\bibfnamefont{Z.}~\bibnamefont{Wang}},
  \bibinfo{author}{\bibfnamefont{S.}~\bibnamefont{Mohamed}}, \bibnamefont{and}
  \bibinfo{author}{\bibfnamefont{N.}~\bibnamefont{De~Freitas}}, in
  \emph{\bibinfo{booktitle}{Proceedings of the 30th International Conference on
  International Conference on Machine Learning - Volume 28}}
  (\bibinfo{publisher}{JMLR.org}, \bibinfo{year}{2013}), ICML'13, pp.
  \bibinfo{pages}{III--1462--III--1470}.

\bibitem[{\citenamefont{Hoffman and Gelman}(2014)}]{Hoffman2014}
\bibinfo{author}{\bibfnamefont{M.~D.} \bibnamefont{Hoffman}} \bibnamefont{and}
  \bibinfo{author}{\bibfnamefont{A.}~\bibnamefont{Gelman}},
  \bibinfo{journal}{J. Mach. Learn. Res.} \textbf{\bibinfo{volume}{15}},
  \bibinfo{pages}{1593} (\bibinfo{year}{2014}).

\bibitem[{\citenamefont{Hellman et~al.}(2013)\citenamefont{Hellman, Steneteg,
  Abrikosov, and Simak}}]{Hellman2013}
\bibinfo{author}{\bibfnamefont{O.}~\bibnamefont{Hellman}},
  \bibinfo{author}{\bibfnamefont{P.}~\bibnamefont{Steneteg}},
  \bibinfo{author}{\bibfnamefont{I.~A.} \bibnamefont{Abrikosov}},
  \bibnamefont{and} \bibinfo{author}{\bibfnamefont{S.~I.} \bibnamefont{Simak}},
  \bibinfo{journal}{Phys. Rev. B} \textbf{\bibinfo{volume}{87}},
  \bibinfo{pages}{104111} (\bibinfo{year}{2013}).

\bibitem[{\citenamefont{Grabowski et~al.}(2009)\citenamefont{Grabowski, Ismer,
  Hickel, and Neugebauer}}]{Grabowski2009}
\bibinfo{author}{\bibfnamefont{B.}~\bibnamefont{Grabowski}},
  \bibinfo{author}{\bibfnamefont{L.}~\bibnamefont{Ismer}},
  \bibinfo{author}{\bibfnamefont{T.}~\bibnamefont{Hickel}}, \bibnamefont{and}
  \bibinfo{author}{\bibfnamefont{J.}~\bibnamefont{Neugebauer}},
  \bibinfo{journal}{Phys. Rev. B} \textbf{\bibinfo{volume}{79}},
  \bibinfo{pages}{134106} (\bibinfo{year}{2009}).

\bibitem[{\citenamefont{Duff et~al.}(2015)\citenamefont{Duff, Davey,
  Korbmacher, Glensk, Grabowski, Neugebauer, and Finnis}}]{Duff2015}
\bibinfo{author}{\bibfnamefont{A.~I.} \bibnamefont{Duff}},
  \bibinfo{author}{\bibfnamefont{T.}~\bibnamefont{Davey}},
  \bibinfo{author}{\bibfnamefont{D.}~\bibnamefont{Korbmacher}},
  \bibinfo{author}{\bibfnamefont{A.}~\bibnamefont{Glensk}},
  \bibinfo{author}{\bibfnamefont{B.}~\bibnamefont{Grabowski}},
  \bibinfo{author}{\bibfnamefont{J.}~\bibnamefont{Neugebauer}},
  \bibnamefont{and} \bibinfo{author}{\bibfnamefont{M.~W.}
  \bibnamefont{Finnis}}, \bibinfo{journal}{Phys. Rev. B}
  \textbf{\bibinfo{volume}{91}}, \bibinfo{pages}{214311}
  (\bibinfo{year}{2015}).

\bibitem[{\citenamefont{Drautz and F\"ahnle}(2004)}]{Drautz2004}
\bibinfo{author}{\bibfnamefont{R.}~\bibnamefont{Drautz}} \bibnamefont{and}
  \bibinfo{author}{\bibfnamefont{M.}~\bibnamefont{F\"ahnle}},
  \bibinfo{journal}{Phys. Rev. B} \textbf{\bibinfo{volume}{69}},
  \bibinfo{pages}{104404} (\bibinfo{year}{2004}).

\bibitem[{\citenamefont{Drautz and F\"ahnle}(2005)}]{Drautz2005}
\bibinfo{author}{\bibfnamefont{R.}~\bibnamefont{Drautz}} \bibnamefont{and}
  \bibinfo{author}{\bibfnamefont{M.}~\bibnamefont{F\"ahnle}},
  \bibinfo{journal}{Phys. Rev. B} \textbf{\bibinfo{volume}{72}},
  \bibinfo{pages}{212405} (\bibinfo{year}{2005}).

\bibitem[{\citenamefont{Mrovec et~al.}(2011)\citenamefont{Mrovec, Nguyen-Manh,
  Els\"asser, and Gumbsch}}]{Matous2011}
\bibinfo{author}{\bibfnamefont{M.}~\bibnamefont{Mrovec}},
  \bibinfo{author}{\bibfnamefont{D.}~\bibnamefont{Nguyen-Manh}},
  \bibinfo{author}{\bibfnamefont{C.}~\bibnamefont{Els\"asser}},
  \bibnamefont{and} \bibinfo{author}{\bibfnamefont{P.}~\bibnamefont{Gumbsch}},
  \bibinfo{journal}{Phys. Rev. Lett.} \textbf{\bibinfo{volume}{106}},
  \bibinfo{pages}{246402} (\bibinfo{year}{2011}).

\bibitem[{\citenamefont{Antropov et~al.}(1996)\citenamefont{Antropov,
  Katsnelson, Harmon, van Schilfgaarde, and Kusnezov}}]{Antropov1996}
\bibinfo{author}{\bibfnamefont{V.~P.} \bibnamefont{Antropov}},
  \bibinfo{author}{\bibfnamefont{M.~I.} \bibnamefont{Katsnelson}},
  \bibinfo{author}{\bibfnamefont{B.~N.} \bibnamefont{Harmon}},
  \bibinfo{author}{\bibfnamefont{M.}~\bibnamefont{van Schilfgaarde}},
  \bibnamefont{and} \bibinfo{author}{\bibfnamefont{D.}~\bibnamefont{Kusnezov}},
  \bibinfo{journal}{Phys. Rev. B} \textbf{\bibinfo{volume}{54}},
  \bibinfo{pages}{1019} (\bibinfo{year}{1996}).

\bibitem[{\citenamefont{Gilbert}(2004)}]{Gilbert2004}
\bibinfo{author}{\bibfnamefont{T.~L.} \bibnamefont{Gilbert}},
  \bibinfo{journal}{IEEE Transactions on Magnetics}
  \textbf{\bibinfo{volume}{40}}, \bibinfo{pages}{3443} (\bibinfo{year}{2004}).

\bibitem[{\citenamefont{McLachlan et~al.}(2014)\citenamefont{McLachlan, Modin,
  and Verdier}}]{McLachlan2014}
\bibinfo{author}{\bibfnamefont{R.~I.} \bibnamefont{McLachlan}},
  \bibinfo{author}{\bibfnamefont{K.}~\bibnamefont{Modin}}, \bibnamefont{and}
  \bibinfo{author}{\bibfnamefont{O.}~\bibnamefont{Verdier}},
  \bibinfo{journal}{Phys. Rev. E} \textbf{\bibinfo{volume}{89}},
  \bibinfo{pages}{061301} (\bibinfo{year}{2014}).

\bibitem[{\citenamefont{Krech et~al.}(1998)\citenamefont{Krech, Bunker, and
  Landau}}]{KRECH19981}
\bibinfo{author}{\bibfnamefont{M.}~\bibnamefont{Krech}},
  \bibinfo{author}{\bibfnamefont{A.}~\bibnamefont{Bunker}}, \bibnamefont{and}
  \bibinfo{author}{\bibfnamefont{D.}~\bibnamefont{Landau}},
  \bibinfo{journal}{Computer Physics Communications}
  \textbf{\bibinfo{volume}{111}}, \bibinfo{pages}{1 } (\bibinfo{year}{1998}).

\bibitem[{\citenamefont{Swope et~al.}(1982)\citenamefont{Swope, Andersen,
  Berens, and Wilson}}]{Swope1982}
\bibinfo{author}{\bibfnamefont{W.~C.} \bibnamefont{Swope}},
  \bibinfo{author}{\bibfnamefont{H.~C.} \bibnamefont{Andersen}},
  \bibinfo{author}{\bibfnamefont{P.~H.} \bibnamefont{Berens}},
  \bibnamefont{and} \bibinfo{author}{\bibfnamefont{K.~R.}
  \bibnamefont{Wilson}}, \bibinfo{journal}{The Journal of Chemical Physics}
  \textbf{\bibinfo{volume}{76}}, \bibinfo{pages}{637} (\bibinfo{year}{1982}).

\bibitem[{\citenamefont{Omelyan et~al.}(2001)\citenamefont{Omelyan, Mryglod,
  and Folk}}]{Omelyan2001}
\bibinfo{author}{\bibfnamefont{I.~P.} \bibnamefont{Omelyan}},
  \bibinfo{author}{\bibfnamefont{I.~M.} \bibnamefont{Mryglod}},
  \bibnamefont{and} \bibinfo{author}{\bibfnamefont{R.}~\bibnamefont{Folk}},
  \bibinfo{journal}{Phys. Rev. Lett.} \textbf{\bibinfo{volume}{86}},
  \bibinfo{pages}{898} (\bibinfo{year}{2001}).

\bibitem[{\citenamefont{Rosengaard and Johansson}(1997)}]{Rosengaard1997}
\bibinfo{author}{\bibfnamefont{N.~M.} \bibnamefont{Rosengaard}}
  \bibnamefont{and}
  \bibinfo{author}{\bibfnamefont{B.}~\bibnamefont{Johansson}},
  \bibinfo{journal}{Phys. Rev. B} \textbf{\bibinfo{volume}{55}},
  \bibinfo{pages}{14975} (\bibinfo{year}{1997}).

\bibitem[{\citenamefont{Suzuki}(1977)}]{Suzuki1977}
\bibinfo{author}{\bibfnamefont{M.}~\bibnamefont{Suzuki}},
  \bibinfo{journal}{Progress of Theoretical Physics}
  \textbf{\bibinfo{volume}{58}}, \bibinfo{pages}{1142} (\bibinfo{year}{1977}).

\bibitem[{\citenamefont{Holm and Janke}(1993)}]{Holm1993}
\bibinfo{author}{\bibfnamefont{C.}~\bibnamefont{Holm}} \bibnamefont{and}
  \bibinfo{author}{\bibfnamefont{W.}~\bibnamefont{Janke}},
  \bibinfo{journal}{Phys. Rev. B} \textbf{\bibinfo{volume}{48}},
  \bibinfo{pages}{936} (\bibinfo{year}{1993}).

\bibitem[{\citenamefont{Hammerschmidt et~al.}(2019)\citenamefont{Hammerschmidt,
  Seiser, Ford, Ladines, Schreiber, Wang, Jenke, Lysogorskiy, Teijeiro, Mrovec
  et~al.}}]{Hammerschmidt2018}
\bibinfo{author}{\bibfnamefont{T.}~\bibnamefont{Hammerschmidt}},
  \bibinfo{author}{\bibfnamefont{B.}~\bibnamefont{Seiser}},
  \bibinfo{author}{\bibfnamefont{M.}~\bibnamefont{Ford}},
  \bibinfo{author}{\bibfnamefont{A.}~\bibnamefont{Ladines}},
  \bibinfo{author}{\bibfnamefont{S.}~\bibnamefont{Schreiber}},
  \bibinfo{author}{\bibfnamefont{N.}~\bibnamefont{Wang}},
  \bibinfo{author}{\bibfnamefont{J.}~\bibnamefont{Jenke}},
  \bibinfo{author}{\bibfnamefont{Y.}~\bibnamefont{Lysogorskiy}},
  \bibinfo{author}{\bibfnamefont{C.}~\bibnamefont{Teijeiro}},
  \bibinfo{author}{\bibfnamefont{M.}~\bibnamefont{Mrovec}},
  \bibnamefont{et~al.}, \bibinfo{journal}{Computer Physics Communications}
  \textbf{\bibinfo{volume}{235}}, \bibinfo{pages}{221 } (\bibinfo{year}{2019}).

\bibitem[{\citenamefont{Ambegaokar and Troyer}(2010)}]{Ambegaokar2009}
\bibinfo{author}{\bibfnamefont{V.}~\bibnamefont{Ambegaokar}} \bibnamefont{and}
  \bibinfo{author}{\bibfnamefont{M.}~\bibnamefont{Troyer}},
  \bibinfo{journal}{American Journal of Physics} \textbf{\bibinfo{volume}{78}},
  \bibinfo{pages}{150} (\bibinfo{year}{2010}).

\bibitem[{Vsi()}]{Vsim}
\emph{\bibinfo{title}{http://inac.cea.fr/l\_sim/v\_sim/}}.

\bibitem[{\citenamefont{Ma and Dudarev}(2012{\natexlab{b}})}]{Ma2012}
\bibinfo{author}{\bibfnamefont{P.-W.} \bibnamefont{Ma}} \bibnamefont{and}
  \bibinfo{author}{\bibfnamefont{S.~L.} \bibnamefont{Dudarev}},
  \bibinfo{journal}{Phys. Rev. B} \textbf{\bibinfo{volume}{86}},
  \bibinfo{pages}{054416} (\bibinfo{year}{2012}{\natexlab{b}}).

\bibitem[{\citenamefont{K\"ormann
  et~al.}(2011{\natexlab{a}})\citenamefont{K\"ormann, Dick, Hickel, and
  Neugebauer}}]{Fritz2011}
\bibinfo{author}{\bibfnamefont{F.}~\bibnamefont{K\"ormann}},
  \bibinfo{author}{\bibfnamefont{A.}~\bibnamefont{Dick}},
  \bibinfo{author}{\bibfnamefont{T.}~\bibnamefont{Hickel}}, \bibnamefont{and}
  \bibinfo{author}{\bibfnamefont{J.}~\bibnamefont{Neugebauer}},
  \bibinfo{journal}{Phys. Rev. B} \textbf{\bibinfo{volume}{83}},
  \bibinfo{pages}{165114} (\bibinfo{year}{2011}{\natexlab{a}}).

\bibitem[{\citenamefont{Ruban and Peil}(2018)}]{Ruban2018}
\bibinfo{author}{\bibfnamefont{A.~V.} \bibnamefont{Ruban}} \bibnamefont{and}
  \bibinfo{author}{\bibfnamefont{O.~E.} \bibnamefont{Peil}},
  \bibinfo{journal}{Phys. Rev. B} \textbf{\bibinfo{volume}{97}},
  \bibinfo{pages}{174426} (\bibinfo{year}{2018}).

\bibitem[{\citenamefont{K\"ormann
  et~al.}(2011{\natexlab{b}})\citenamefont{K\"ormann, Dick, Hickel, and
  Neugebauer}}]{Kormann2011}
\bibinfo{author}{\bibfnamefont{F.}~\bibnamefont{K\"ormann}},
  \bibinfo{author}{\bibfnamefont{A.}~\bibnamefont{Dick}},
  \bibinfo{author}{\bibfnamefont{T.}~\bibnamefont{Hickel}}, \bibnamefont{and}
  \bibinfo{author}{\bibfnamefont{J.}~\bibnamefont{Neugebauer}},
  \bibinfo{journal}{Phys. Rev. B} \textbf{\bibinfo{volume}{83}},
  \bibinfo{pages}{165114} (\bibinfo{year}{2011}{\natexlab{b}}).

\end{thebibliography}
\bibliographystyle{apsrev}

\end{document}